\documentclass[prd,floatfix,showpacs]{revtex4}
\usepackage{amsthm}
\usepackage{amsmath}
\usepackage{amsfonts}
\usepackage{graphicx,epsfig}

\newcommand{\bk}{\mathbf{k}}

\def\slashi#1{\rlap{\sl/}#1}
\def\slashii#1{\setbox0=\hbox{$#1$}             
   \dimen0=\wd0                                 
   \setbox1=\hbox{\sl/} \dimen1=\wd1            
   \ifdim\dimen0>\dimen1                        
      \rlap{\hbox to \dimen0{\hfil\sl/\hfil}}   
      #1                                        
   \else                                        
      \rlap{\hbox to \dimen1{\hfil$#1$\hfil}}   
      \hbox{\sl/}                               
   \fi}                                         %
\def\slashiii#1{\setbox0=\hbox{$#1$}#1\hskip-\wd0\hbox to\wd0{\hss\sl/\/\hss}}
\newcommand{\nn}{\nonumber}

\newcommand{\ba}{\begin{eqnarray}}
\newcommand{\ea}{\end{eqnarray}}

\newtheorem{theorem}{Theorem}

\reversemarginpar

\newcommand{\NI}{I}
\newcommand{\NV}{V}
\newcommand{\NL}{L}
\newcommand{\gammaE}{\gamma_{E}}
\newcommand{\opS}{\hat{\mathcal{S}}}
\newcommand{\opO}{\hat{\mathcal{O}}}
\newcommand{\opA}{\hat{\mathcal{A}}}
\newcommand{\Tbar}{\cal\bar T}

\begin{document}
\title{The thermal operator representation for Matsubara sums}

\author{Olivier Espinosa}
\email{olivier.espinosa@usm.cl}
\affiliation{Departamento de F{\'\i}sica, Universidad T{\'e}cnica Federico
Santa Mar{\'\i}a, Casilla 110-V, Valpara{\'\i}so, Chile}
\begin{abstract}
We prove in full generality the thermal operator representation
for Matsubara sums in a relativistic field theory of scalar and
fermionic particles. It states that the full result of performing
the Matsubara sum associated to any given Feynman graph, in the
imaginary-time formalism of finite-temperature field theory, can
be directly obtained from its corresponding zero-temperature energy
integral, by means of a simple linear operator, which is
independent of the external Euclidean energies and whose form
depends solely on the topology of the graph.
\end{abstract}

\maketitle

\section{Introduction}
\label{introduction}

Relativistic quantum field theory is the mathematical formalism that allows
us to describe the interactions among elementary particles, according
to the principles of quantum mechanics and special relativity.
When considering the physics of very dense or very hot plasmas,
such as the early universe or the quark-gluon plasma that should be produced in
heavy ion collisions, it
becomes necessary to formulate the relevant questions in terms of thermally
averaged quantities, according to
statistical mechanics. Most of the physical information, for
systems both in and slightly out of thermal equilibrium, can be obtained from
the study of the so-called thermal Green functions, defined as thermal
expectation values of a time-ordered product of field operators.
One fruitful and widely used approach to study these Green functions is
perturbation theory, where these are computed by means of a systematic
expansion in terms of Feynman diagrams\cite{books-on-FTQFT}.

The simplest diagrammatic analysis is obtained in the so-called
imaginary-time formalism, in which the diagrams have the same topology
and are computed according to basically the same Feynman rules
as in the zero-temperature Euclidean theory, except
for one very important difference: the energy component of both the
external 4-momenta and the internal 4-momenta carried by the
propagators is quantized (in slightly different ways, according to
the nature --bosonic or fermionic-- of the associated particle), that
is, it must be a \emph{Matsubara frequency}\cite{matsubara}.
Accordingly, the calculation of a loop diagram in the imaginary-time
formalism of quantum field theory at finite temperature necessarily
involves sums over internal Matsubara frequencies\cite{books-on-FTQFT},
an operation that we shall generically call \emph{the
Matsubara sum} associated with the graph. Although this sum can be
computed in a number of ways, usually in a systematic fashion,
such computations can become quite tedious for higher loop
diagrams\cite{saclay-method,Guerin}. Another related difficulty of the
the imaginary-time formalism is the separation of the vacuum contribution
from the finite temperature corrections, since
the naive zero-temperature limit of any given loop graph is of
the indeterminate type $0\cdot\infty$ \cite{blaizot+reinosa}.

It was discovered some time
ago\cite{espinosa+stockmeyer} that, for some particular classes of
diagrams in a scalar theory, the full result of performing the
Matsubara sum associated with a Feynman graph could be completely
determined from its zero-temperature counterpart, by means of a
simple linear operator that depends on the topology of the diagram
but is independent of the Euclidean energies carried by its
external lines. Due to the simple and well-defined structure of
this operator, dubbed \emph{the thermal operator}, it was
conjectured that this result should hold for all Feynman graphs.

In the recent paper \cite{blaizot+reinosa} it has been partially
shown that this conjecture is actually correct. The authors of
reference \cite{blaizot+reinosa} have rescued from oblivion an
old, systematic and very elegant method to perform the Matsubara
sum associated with an arbitrary graph, due to
M.~Gaudin\cite{gaudin}, and have used it, as an illustration of
the power of Gaudin's method, to show that the main part of
our conjecture follows naturally from it. The proof
given in reference \cite{blaizot+reinosa} does address the
relationship between the structure of the full thermal result and
the zero-temperature result, but stops short of explicitly
constructing the thermal operator that relates them. Also, it
leaves undiscussed conjectures 2 and 3 presented in reference
\cite{espinosa+stockmeyer}.

\medskip

In this letter we restate all the conjectures of
reference\cite{espinosa+stockmeyer} as full-blown theorems
concerning general properties of the Feynman graphs in the
imaginary-time formalism of thermal field theory, extending their
validity to theories containing Dirac fields as well. Following
reference \cite{blaizot+reinosa}, we make full use of Gaudin's
method to prove these theorems, although in a slightly different
and more explicit form.

The structure of the rest of the paper is as follows: In section
\ref{sec:thermal-operator} we present, in the form of theorems, the
general properties of the finite-temperature imaginary-time
Feynman graphs that were previously presented as conjectures,
in terms of the thermal operator, suitably extended to incorporate
possible fermionic lines. In section \ref{sec:gaudin} we summarize
Gaudin's method to perform the sums over Matsubara frequencies,
which will be central to the proofs of the theorems, presented in
sections \ref{sec:proofs-scalar} (scalars) and
\ref{sec:proofs-fermions} (fermions). Our conclusions are
presented in section \ref{sec:conclusions}.

\medskip

\section{The thermal operator}
\label{sec:thermal-operator}

In a scalar field theory, the mathematical expression
corresponding to an amputated graph with $V$ vertices,
$\NI$ internal lines, and external 4-momenta
$P_\alpha=(p_{\alpha}, {\bf p}_\alpha)$ has the form
\begin{equation}
\label{eq:D-def}
\frac{(-\lambda)^{V}}S\int
[\prod\limits_{i=1}^{\NI}\frac{d^3k_i}{\left( 2\pi \right)^3 2E_i}%
\prod\limits_{v=1}^{\NV-1} (2\pi )^3\delta^{(3)} ({\bf k}_v)]
\;D(p, E, T),
\end{equation}
where $\lambda$ represents the coupling constant and $S$ is the
symmetry factor of the graph; ${\bf k}_i$ is the spatial
3-momentum of the $i$-th internal line and $E_i:=({\bf
k}_i^2+m_i^2)^{1/2}$ is its associated kinematic energy; ${\bf
k}_v$ denotes the total 3-momentum entering vertex $v$; the
unsubscripted symbols $p$ and $E$ denote, respectively, the full
set of Euclidean external and kinematic internal energies,
$p:=\left\{p_1, p_2, \dots, p_{\NV}\right\}$ (with $\sum_{i=1}^\NV p_i=0$)
and $E:=\left\{E_1, E_2, \dots, E_\NI\right\}$; and $T$ is the temperature.
The delta functions ensure conservation of spatial 3-momentum at
each vertex, so that the integration measure reduces essentially
to an integration over the 3-momenta of the $\NL=\NI-\NV+1$ independent
loops. In the finite temperature Euclidean formalism all scalar lines,
external and internal, carry discrete Euclidean energies which are
integer multiples of $2\pi T$. Each internal line has an
associated Matsubara frequency, denoted by $k_i = \omega_{n_i} := 2\pi T n_i$.
The $D$-function is given by the normalized Matsubara sum
\begin{equation}
\label{def-D-function}
D(p, E, T)= \gammaE \, T^{\NL}\sum_{\{n_i\}}
\prod_{i=1}^\NI \Delta(k_i, E_i)\delta(p, k),
\end{equation}
where
\ba
\label{eq:def-gamma}
\gammaE:=\prod_{i=1}^{\NI}2E_i,
\ea
$\NL$ is the number of independent loops in the graph, and
$\Delta(k_i,E_i)$ is the scalar propagator associated with the
$i$-th internal line, with
\ba\label{eq:def-Delta}
\Delta(k,E):=\frac{1}{k^2+E^2}.
\ea
The sums over each $n_i$ ($i=1, \dots,\NI$) run from $-\infty$ to $+\infty$. The
$\delta$-function, with $k=\left\{k_1, \dots, k_\NI\right\}$, is a
generalized Kronecker delta which ensures conservation of energy
at each vertex. The topology of the diagram is totally contained
in this generalized delta.
\medskip

In a theory containing both fermions and scalars, the structure of
\eqref{eq:D-def} is basically unchanged, except
for extra spin indices carried by the external fermionic lines and
a possible extra sign associated with fermionic loops.
The $D$-function is still given by \eqref{def-D-function}, except
that each fermionic line carries a Matsubara frequency
consistent with antiperiodic boundary conditions on the fermionic fields,
\begin{equation}
\label{fermionic_Matsubara frequency}
k = \tilde{\omega}_n := 2\pi T \left(n+1/2\right),
\end{equation}
entering through the fermionic propagator, which now has a matrix structure
and depends explicitly on the spatial momentum $\bk$,
\begin{equation}
\label{fermionic_propagator}
\tilde \Delta (k,{\bk}) = \frac{{m + ik\gamma^0- \mathbf{k}\cdot\boldsymbol{\gamma}}}
{{k^2  + E_k^2 }},
\end{equation}
where the $\gamma^\mu=(\gamma^0, \boldsymbol{\gamma})$
are the usual Minkowski space gamma matrices, obeying the standard
anticommutation relations,
$\left\{\gamma^{\mu},\gamma^{\nu}\right\}=2\eta^{\mu\nu}$.

In order to state our general theorems in a way that is applicable to
a general theory containing both scalars and fermions,
it will be convenient to define the following generalized \emph{signed}
thermal occupation number function, which takes into account the
statistics corresponding to both kinds of internal lines:
\begin{equation}
\label{def-N}
N(E):=
\left\{%
\begin{array}{ll}
    n(E), & \hbox{for a scalar line;} \\
    -\tilde n(E), & \hbox{for a fermionic line,} \\
\end{array}%
\right.
\end{equation}
where $n(E) = \left(e^{\beta E}-1\right)^{-1}$ and $\tilde n(E) =
\left(e^{\beta E}+1\right)^{-1}$ are, respectively, the Bose-Einstein and
Fermi-Dirac thermal occupation factors, for the case of vanishing
chemical potential.

Additionally, the theorems will be formulated in terms of a simple
reflection operator on the space of functions of several variables,
\begin{equation}
\label{def-S}
\opS_E f(E,x):=f(-E,x),
\end{equation}
where $E$ stands for any one variable and $x$ for all the others.

\noindent\
The main result presented in this paper is enunciated in the next theorem.
It basically states that the ``energy part'' of any Feynman graph in the
finite-temperature imaginary-time formalism, represented here in terms of
the $D$-function introduced in \eqref{eq:D-def}, can be obtained
directly from the corresponding zero-temperature energy integral.

\bigskip

\begin{theorem}\label{theorem-1}
\noindent\textbf{[Thermal Operator Representation]}
The $D$-function defined in \eqref{def-D-function} for an
amputated Feynman graph can be expressed in the form
\begin{equation}\label{eq:theorem-1}
D(p, E, T)=\opO(E, T)D_0(\omega, E)
\Big|_{\omega=p},
\end{equation}
where $D_0(\omega, E)$ is the $D$-function of the
Euclidean zero-temperature graph and $\opO(E, T)$, the
thermal operator, is the following linear operator:
\begin{equation}\label{eq:thermal-operator-form1}
\begin{split}
\opO(E,T):=1+&\sum_{i=1}^\NI  N_i(1+\opS_i)
+\sideset{}{^\prime}\sum_{\langle i_1,  i_2\rangle} N_{i_1}N_{i_2}
(1+\opS_{i_1})(1+\opS_{i_2})\\
+& \dots + \sideset{}{^\prime}\sum_{\langle i_1, \dots,  i_{\NL}\rangle}
\prod_{l=1}^{\NL} N_{i_l}(1+\opS_{i_l}).
\end{split}
\end{equation}
Here $N_i\equiv N(E_i)$, where
$N(E)$ is the generalized signed thermal occupation factor
defined in \eqref{def-N}; $\opS_i:=\opS_{E_i}$, where $\opS_E$ is the reflection
operator defined in \eqref{def-S}; the indices $i_1, i_2, \ldots$ run
from 1 to $\NI$ (the number of internal propagators) and the symbol
$\langle i_1, \dots,  i_k\rangle$ stands for an unordered
$k$-tuple with no repeated indices, representing a particular set
of internal lines. The primes on the summation symbols imply that
certain tuples $\langle i_1, \dots, i_k\rangle$ are to be excluded
from the sums: those such that if we snip all the lines
$i_1,\dots,i_k$ then the graph becomes disconnected.
\end{theorem}

Note that the operator $\opO(E,T)$ contains products of at most
$\NL$ thermal occupation factors $N(E_i)$, since for a $\NL$-loop
graph the maximum number of lines that can be snipped without
disconnecting the graph is precisely $\NL$. This generic feature
of the thermal graph in the imaginary-time formalism is of course
well known. However, based on a general property of the
zero-temperature $D$-function, formulated in theorem
\ref{theorem-3} below, it is also possible to use a modified
thermal operator, which has a simpler algebraic form:

\medskip

\begin{theorem}\label{theorem-2}
\noindent\textbf{[Simpler form of the Thermal Operator]}
When acting on the zero-temperature $D$-function, $D_0(p,
E)$, the thermal operator $\opO(E, T)$ can be
replaced by the the simpler
\begin{equation}\label{eq:thermal-operator-form2}
\opO_\star(E, T)=\prod_{i=1}^\NI [1+N_i(1+\opS_i)].
\end{equation}
\end{theorem}

Note that the operator $\opO_\star(E, T)$ in
\eqref{eq:thermal-operator-form2} can be expanded as in
\eqref{eq:thermal-operator-form1}, with the only difference that the summation
symbols will carry no primes, that is, all tuples $\langle i_1, \dots,
i_k\rangle$ ($1\le k \le \NI$) will be allowed in the sum.
Clearly, forms  \eqref{eq:thermal-operator-form1}
and \eqref{eq:thermal-operator-form2} of the thermal operator
will be equivalent if we can prove that tuples associated with
disconnected graphs (the ones excluded from the summations in
\eqref{eq:thermal-operator-form1}) give rise to operators that
produce a vanishing contribution to the $D$-function in
\eqref{eq:theorem-1}. This is the content of our last theorem:
\bigskip

\begin{theorem}\label{theorem-3}
\noindent\textbf{[Cut sets do not contribute]}
The zero-temperature $D$-function,
$D_0(\omega, E)$, is annihilated by the operators
\begin{equation}\label{annihilation-operator}
\opA(C):=\prod_{i_l\in C}(1+\opS_{i_l}),
\end{equation}
where $C$ stands for a \emph{cut set} of the graph, that is,
any set of indices $i_1,\ldots,i_k$ such that the graph becomes
disconnected if the corresponding lines are snipped.
\end{theorem}
\medskip

The concept of \emph{cut set}, as used here, bears no connection to the
concepts of cut and cut diagrams as they are usually understood
in diagrammatic quantum field theory. Cut sets are determined solely
by the topology of the diagram, and have no further mathematical
or physical meaning.
\medskip

\section{Gaudin method}
\label{sec:gaudin}

Here we present a summary of Gaudin's method and its use in the
computation of the Matsubara $D$-function, in the purely scalar case.
For all the details see references \cite{gaudin}
and \cite{blaizot+reinosa}. The changes in the presence of
fermionic lines are commented upon at the end of this section.

Gaudin's method to perform the Matsubara sum associated to a given graph
is based on two main ideas. The first is to make use of the spectral
representation of the propagator, which puts the dependence on the Matsubara
frequency linearly in the denominator. For the scalar case it reads
\begin{equation}
\label{scalar spectral representation}
\Delta (\omega _n ,{\mathbf{k}}) = \int_{ - \infty }^\infty  {\frac{{dk^0 }}
{{2\pi }}} \frac{{\rho \left( {k^0 ,{\mathbf{k}}} \right)}}
{{k^0  - i\omega _n }},
\end{equation}
with the spectral function given by
\ba
\label{scalar spectral function}
\rho \left( {k^0 ,{\mathbf{k}}} \right) &=& 2\pi \epsilon (k^0 )\delta (k^2  - m^2 )
\\
& = &\frac{{2\pi }}
{{2E_k }}\left[ {\delta (k^0  - E_k ) - \delta (k^0  + E_k )} \right].
\ea
$\epsilon (k^0 )$ is the sign of $k^0$.
\medskip

Once all the propagators have been represented by means of
\eqref{scalar spectral representation}, the Matsubara sum to be
performed takes the form
\begin{equation}
\label{Matsubara sum Gaudin}
T^{\NL}\sum_{\{n_i\}}
\prod_{i=1}^\NI \frac{1}{k_i^0  - ik_i }\delta(p, k),
\end{equation}
where the $k_i=\omega_{n_i}$ are the Matsubara frequencies and
the $k_i^0$ are external real variables. The generalized Kronecker
delta $\delta(p,k)$ enforces $\NV-1$ independent linear relations
satisfied by the Matsubara frequencies $k$, also involving the
external Euclidean energies $p$, which we shall write (following
reference \cite{blaizot+reinosa}) as
\ba
\label{R:vertex constraints}
R_v (p,k) = 0,\quad\text{for }v = 1, \ldots ,\NV - 1.
\ea
This system of linear equations allows us to solve for $\NV - 1$
of the $\NI$ Matsubara frequencies in terms of a set of $\NL=\NI-\NV+1$
independent ones. In general, there will be several distinct ways of
choosing this set of independent Matsubara frequencies.
As shown by Gaudin, there is a one-to-one correspondence between
the collection of all possible sets of independent Matsubara frequencies and
the set of all \emph{trees} associated to the given (connected) diagram
$\Gamma$.

A tree is a set of lines of $\Gamma$ joining all vertices and making a connected
graph with no loops. Every tree $\cal T$ will contain $\NV-1$ lines and its
complement $\Tbar$ (the set of lines which do not belong to $\cal T$) will
have $\NL$ lines. The Matsubara frequencies corresponding to the $\NL$ lines in
$\Tbar$, denoted by $k_l$, will constitute a set of independent Matsubara
frequencies in terms of which the system \eqref{R:vertex constraints} can be solved.
The Matsubara frequencies
associated with the lines of the tree, $k_j$, with $j\in\cal T$,
will be linear combinations of the independent Matsubara frequencies
and the external Euclidean energy variables,
\ba
k_j=\Omega_j^{\cal T}(p,k_l),\quad j\in{\cal T}, l\in{\Tbar}.
\ea
\begin{figure}
\centerline{\epsfig{file=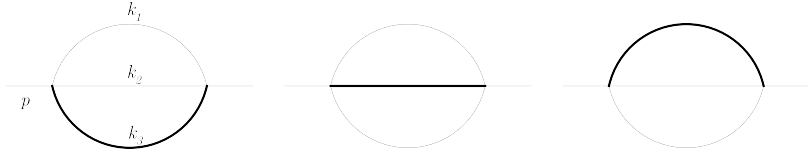,height=2cm,angle=0}}
\caption[]{The trees of a simple two-loop graph, shown as dark lines.}
\label{fig:trees}
\end{figure}

As a simple example, in figure \ref{fig:trees} we show the three
possible trees for the two-loop two-vertex graph shown.
In this case, each tree $\cal T$ is composed by a single internal line
(heavy line), whose Matsubara frequency can be expressed, after using energy
conservation at one of the vertices, in terms
of the two independent Matsubara frequencies associated with the
(thin) lines that do not belong to the tree (these are the lines in $\Tbar$) and the external
energy $p$. For instance, for the first tree we have
$k_3 = p - k_1 - k_2$, etc.

Gaudin's main insight is the following identity for the rational function
appearing in \eqref{Matsubara sum Gaudin}:
\ba\label{eq:Gaudin-decomposition}
\sum_{\{n_i\}}\prod\limits_{i = 1}^\NI {\frac{1}{{k_i^0  - ik_i }}}\delta(p, k)
= \sum\limits_{\cal T} {\prod\limits_{j \in \cal T} {\frac{1}
{{k_j^0  - i\Omega _j^{\cal T} (p, - ik_l^0 )}}} }
\sum_{\{n_l\}}\prod\limits_{l \in\Tbar} {\frac{1}{{k_l^0  - ik_l}}}.
\ea
In our example in figure \ref{fig:trees}, this identity takes the
form
\begin{multline*}
\sum\limits_{n_1 ,n_2 ,n_3 } {\frac{{\delta \left( {p - k_1  - k_2  - k_3 } \right)}}
{{\left( {k_1^0  - ik_1 } \right)\left( {k_2^0  - ik_2 } \right)\left( {k_3^0  - ik_3 } \right)}}}  = \frac{1}
{{\left( {k_1^0  + k_2^0  + k_3^0  - ip} \right)}}
\\
\times \left\{ {\sum\limits_{n_1 ,n_2 } {\frac{1}
{{\left( {k_1^0  - ik_1 } \right)\left( {k_2^0  - ik_2 } \right)}}}  + \sum\limits_{n_1 ,n_3 } {\frac{1}
{{\left( {k_1^0  - ik_1 } \right)\left( {k_3^0  - ik_3 } \right)}}}  + \sum\limits_{n_2 ,n_3 } {\frac{1}
{{\left( {k_2^0  - ik_2 } \right)\left( {k_3^0  - ik_3 } \right)}}} }\right\}.
\end{multline*}
So, using Gaudin's decomposition \eqref{eq:Gaudin-decomposition},
for any given graph the full Matsubara sum is split into a number
of other sums, one for each tree $\cal T$, where each one of these
sums is actually simply a product of independent sums over each of
the Matsubara frequencies in $\Tbar$.
Each independent sum has to be regulated. Gaudin assigns
to each internal line a regulator $e^{ik_i\tau_i}$, where the time $\tau_i$
is taken to zero at the end.
The sum that results is well defined,
\ba
\label{bosonic-sum}
T\sum\limits_{n_l } {\frac{{e^{i\omega _l T_l } }}
{{k_l^0  - ik_l }}}  = \epsilon_l n(\epsilon _l k_l^0 )e^{k_l^0 T_l},
\ea
where $n(k^0)$ is the Bose-Einstein occupation factor.
$T_l$ is some linear combination of the times $\tau_i$ associated to the
lines that belong to the loop defined by $l\in\Tbar$, and $\epsilon_l$
is the sign of $T_l$. Only $\epsilon_l$ matters when the regulators are
removed.

Consequently, Gaudin's result for the $D$-function in the purely scalar case is
\ba
\label{D-Gaudin-result}
D(p,E,T) = \prod\limits_{i = 1}^\NI
\int_{ - \infty }^\infty  dk_i^0 \bar \rho (k_i^0 ,E_i )
\sum\limits_{\cal T}  {\left( {\prod\limits_{j \in {\cal T} }
{\frac{1}{{k_j^0  - i\Omega _j^{\cal T} }}
\prod\limits_{l \in \Tbar } {\epsilon_l n(\epsilon _l k_l^0) } } } \right)},
\ea
where $\Omega_j^{\cal T}=\Omega_j^{\cal T}(p,-i k_l^0)$ and
$\bar\rho$ is the reduced spectral function,
\ba
\label{reduced spectral function}
\bar \rho (k^0 ,E_k )
= \frac{2E_k}{2\pi}\rho \left( {k^0 ,{\mathbf{k}}} \right)
= \delta (k^0  - E_k ) - \delta (k^0  + E_k ).
\ea
\medskip

Gaudin's method applies equally well in
the case the graph contains fermionic lines.
As for the scalar propagator, the fermionic propagator \eqref{fermionic_propagator}
admits a spectral representation,
\ba
\label{fermionic propagator: spectral rep}
\tilde \Delta (\tilde \omega _n ,{\mathbf{k}}) = \int_{ - \infty }^\infty  {\frac{{dk_0 }}
{{2\pi }}} \frac{{\tilde \rho \left( {k_0 ,{\mathbf{k}}} \right)}}
{{k_0  - i\tilde \omega _n }},
\ea
with
\ba
\label{fermionic propagator: spectral function}
\tilde \rho \left( {k_0 ,{\mathbf{k}}} \right) = 2\pi \varepsilon (k_0 )
\left( {\slashi{k} + m} \right)\delta (k^2  - m^2 ),
\ea
where the metric signature is that of Minkowski space,
$\slashi{k} = k_{\mu}\gamma^\mu = k_0\gamma^0-\mathbf{k}\cdot\mathbf{\gamma}$
and $k^2 = k_0^2-\mathbf{k}^2$.
Note that the spectral representation of the fermionic propagator
\eqref{fermionic_propagator} ``hides''
the dependence on the Matsubara frequency $k$ appearing in its numerator,
so that the only difference with the scalar case is the extra matrix
structure $\left( {\slashi{k} + m} \right)$.

Gaudin's method to perform the Matsubara sum in the form \eqref{Matsubara sum Gaudin},
through the decomposition in terms of trees, makes no reference to the bosonic
or fermionic nature of the frequencies to be summed over.
If an independent frequency $\tilde k_l$ ($l\in\bar{\mathcal T}$) is fermionic,
we shall need the sum
\ba
\label{fermionic-sum}
T\sum\limits_{n_l } {\frac{{e^{i\tilde k_l T_l } }}
{{k_l^0  - i\tilde k_l }}}  =  - \epsilon _l \tilde n(\epsilon _l k_l^0 )e^{k_l^0 T_l },
\ea
which can be obtained from the bosonic analog by writing
$k_l^0  - i\tilde k_l  = \left( {k_l^0  - i\pi T} \right) - ik_l$.
$\tilde n(k^0)$ is the Fermi-Dirac function,
\ba
\tilde n(k^0 ) = \frac{1}{{e^{\beta k^0 }  + 1}}.
\ea

We see that the fermionic result \eqref{fermionic-sum} can be simply obtained from
its bosonic analog \eqref{bosonic-sum} through the replacement
\[
n \to  - \tilde n.
\]
Therefore, the result for the $D$-function in the presence of fermionic propagators
will be identical in form to \eqref{D-Gaudin-result}, except that for the fermionic
lines the reduced spectral function $\bar\rho$ will have an extra matrix factor,
and the thermal factor $\epsilon_l n(\epsilon_l k_l^0)$ will be replaced by
$-\epsilon_l \tilde n(\epsilon_l k_l^0)$.

\section{Proofs of the theorems for the scalar case}
\label{sec:proofs-scalar}

In order to prove the theorems enunciated in section \ref{sec:thermal-operator},
we shall need to identify the vacuum ($T=0$) limit of Gaudin's result
\eqref{D-Gaudin-result} for  the Matsubara $D$-function $D(p,E,T)$.
The temperature dependence is solely contained in the thermal factors
$n(\epsilon _l k_l^0)$.
Following reference \cite{blaizot+reinosa}, we use the identity
\ba
\label{BE-identity}
n(k^0 ) =  - \theta ( - k^0 ) + \epsilon (k^0 )n(\left| {k^0 }\right|),
\ea
from which it follows that
\[
\epsilon _l n(\epsilon _l k_l^0)  =  - \epsilon _l \theta ( - \epsilon _l k_l^0 )
+ \epsilon (k_l^0 )n(\left| {k_l^0 } \right|).
\]
Writing the reduced spectral function defined in \eqref{reduced
spectral function} in terms of the reflection operator $\opS_E$ as
\ba
\label{reduced spectral function representation}
\bar \rho (k^0 ,E) = \left( {1 - \opS_E } \right)\delta (k^0  - E),
\ea
and using the notation $\opS_l\equiv\opS_{E_l}$, we can
perform the integrations over each of the variables $k_l^0$ ($l\in\Tbar$)
in \eqref{D-Gaudin-result} as
\ba
\int_{ - \infty }^\infty  {dk_l^0 \bar \rho (} k_l^0 ,E_l )\epsilon _l n(\epsilon _l k_l^0) f(k_l^0 )
&=& \left( {1 - \opS_l } \right)\int_{ - \infty }^\infty  {dk_l^0 \delta (} k_l^0  - E_l )
\epsilon _l n(\epsilon _l k_l^0) f(k_l^0 )
\nn\\
&=&\left( {1 - \opS_l } \right)\epsilon _l n(\epsilon _l E_l) f(E_l )
\nn\\
&=&\left( {1 - \opS_l } \right)\left[ { - \epsilon _l \theta ( - \epsilon _l E_l )
+ \epsilon (E_l )n(\left| {E_l } \right|)} \right]f(E_l ),
\nn
\ea
where $E_l$ is for the moment considered as an arbitrary real variable. We now apply the
reflection operator $\opS_l$ explicitly and then use the fact that $E_l$ is actually
a positive quantity. The vacuum part is given by
\ba
\left( {1 - \opS_l } \right)\left[ { - \epsilon _l \theta ( - \epsilon _l E_l )} f(E_l )\right]
&=&  - \epsilon _l \theta ( - \epsilon _l E_l )f(E_l ) + \epsilon _l \theta (\epsilon _l E_l )f( - E_l )
\nn\\
& =& f( - \epsilon _l E_l ),
\nn
\ea
whereas the thermal part is given by
\ba
\left( {1 - \opS_l } \right)\left[ {\epsilon (E_l )n(\left| {E_l } \right|)f(E_l )} \right]
&=& \epsilon (E_l )n(\left| {E_l } \right|)f(E_l ) - \epsilon ( - E_l )n(\left| { - E_l } \right|)f( - E_l )
\nn\\
&=& n_l f(E_l ) + n_l f( - E_l )
\nn\\
&=& n_l \left( {1 + \opS_l } \right)f(E_l )
\nn\\
&=& n_l \left( {1 + \opS_l } \right)f(-\epsilon _l E_l ),
\nn
\ea
since $\left( {1 + \opS_l } \right)f(E_l )=\left( {1 + \opS_l } \right)f(-E_l )$, and
where we have denoted $n_l\equiv n(E_l)$.
\medskip

Therefore,
\ba
\label{D-Espinosa-Gaudin-result}
D(p,E,T) = \sum\limits_{\cal T}  {\prod\limits_{l \in \Tbar }
{\left[ {1 + n_l \left( {1 + \opS_l } \right)} \right]} } D_0^{\cal T} (p,E)
\ea
where $D_0^{\cal T} (p,E)$ is the contribution to the vacuum
$D$-function associated with the tree $\cal T$:
\ba
D_0^{\cal T} (p,E) &=& \prod\limits_{j \in {\cal T} }
{\int_{ - \infty }^\infty  {dp_j^0 \bar \rho (k_j^0 ,E_j )}
\frac{1}{{k_j^0  - i\Omega _j^{\cal T} (p,i\epsilon _l E_l )}}}
\nn\\
&=&
\label{D0-tau-result}
\prod\limits_{j \in {\cal T} } {\left( {1 - \opS_j } \right)\frac{1}
{{E_j  - i\Omega _j^{\cal T} (p,i\epsilon _l E_l )}}}.
\ea
\noindent
Equations \eqref{D-Espinosa-Gaudin-result} and \eqref{D0-tau-result} are
our starting points for the proofs of our three theorems.
To start with, we notice that the function $D_0^{\cal T} (p,E)$ is
annihilated by each of the operators $(1+\opS_j)$ with $j \in {\cal T}$, due
to the identity
\ba
\label{opS-identity}
\left( {1 + \opS_E } \right)\left( {1 - \opS_E } \right) = 1 - \opS_E^2  = 0.
\ea
This allows us to extend the index of the product $\prod\limits_{l \in \bar {\cal T}}$
in \eqref{D-Espinosa-Gaudin-result} to all possible values, for any tree ${\cal T}$:
\[
\prod\limits_{l \in \Tbar } {\left[ {1 + n_l \left( {1 + \opS_l } \right)} \right]} D_0^{\cal T} (p,E)
= \prod\limits_{j = 1}^\NI {\left[ {1 + n_j \left( {1 + \opS_j } \right)} \right]D_0^{\cal T}(p,E)}.
\]
In this form, the operator acting on $D_0^{\cal T}(p,E)$ becomes
$\cal T$-independent, which allows us to move the sum over all trees in
\eqref{D-Espinosa-Gaudin-result} through it:
\ba
D(p,E,T) &=& \prod\limits_{j = 1}^\NI \left[ {1 + n_j \left( {1 + \opS_j } \right)} \right]
\sum\limits_{\cal T} {D_0^{\cal T} (p,E)}
\nn\\
&=& \hat O_* (E,T)D_0 (p,E),
\nn
\ea
and this proves theorem 2.

Theorem 3 is readily proven by noticing that if $C$ is a cut set of the graph
$\Gamma$, then the operator
\[
\opA(C) = \prod\limits_{k \in C} {\left( {1 + \opS_k } \right)}
\]
will contain at least one factor ${\left( {1 + \opS_{j_{\cal T}} } \right)}$ with
${j_{\cal T}}\in \cal T$ for \emph{every} tree $\cal T$. As we have already pointed out,
this factor will annihilate the corresponding $D_0^{\cal T} (p,E)$, and therefore
$D_0(p,E)$ will be annihilated by $\opA(C)$. This proves theorem 3.

Finally, as explained in section \ref{sec:thermal-operator}, theorem 1 follows
directly from theorems 2 and 3.

\section{Extension to fermions}
\label{sec:proofs-fermions}

As it was shown in section \ref{sec:gaudin}, in the presence of fermionic lines
the Matsubara $D$-function has essentially the same structure as in the scalar
case. Now we use the identity
\ba
\label{FD-identity}
\tilde n(k^0 ) = \theta ( - k^0 ) + \epsilon (k^0 )\tilde n(\left| {k^0 } \right|),
\ea
which has the same contents as the well-known $\tilde
n(-E)=1-\tilde n(E)$. We note that both identities
\eqref{BE-identity} and \eqref{FD-identity} can be written in
terms of the generalized occupation number function defined by
\eqref{def-N} as
\ba
\label{N-identity}
N(k^0 ) = -\theta ( - k^0 ) + \epsilon (k^0 )N(\left| {k^0 } \right|),
\ea
and therefore the manipulation of section \ref{sec:proofs-scalar} are valid in
general, with $n_i$ replaced by $N_i$.

The only other difference with the scalar case is the spectral function.
The reduced spectral function for fermions is
\ba
\label{fermionic reduced spectral function}
\tilde {\bar \rho} (k^0, {\mathbf{k}}, E_k )
= \frac{2E_k}{2\pi}\tilde \rho \left( {k_0 ,{\mathbf{k}}} \right)
= \left( {\slashi{k} + m} \right)\left[\delta (k^0 - E_k ) - \delta (k^0 + E_k )\right]
= \left( {1 - \opS_{E_k} } \right)\delta (k^0  - E_k)\left( {\slashi{k} + m}\right).
\ea

The propagator for a fermionic line $i$ will be represented in the form
\ba
\tilde \Delta (\tilde \omega _i ,{\mathbf{k}}_i ) &=&
\frac{1}{{2E_i }}(1 - \opS_i )\int_{ - \infty }^\infty  {dk_i^0 \delta (k_i^0  - E_i )
(\slashi{k} + m)\frac{1}{{k_i^0  - i\tilde \omega _i }}},
\ea
which has the necessary structure for the derivations presented in
the previous section to hold. Hence, in the presence of fermionic
lines the Matsubara $D$-function will still have the form
\eqref{D-Espinosa-Gaudin-result}, apart from the change
$n\to -\tilde n$ to account for the Fermi-Dirac statistics
for fermions and some extra structure in the numerator of the
right-hand side of \eqref{D0-tau-result}. Therefore, the proofs
presented at the end of the previous section are unaffected.

\section{Conclusions}
\label{sec:conclusions}

In this paper we have proven rigorously, extending the approach
presented in reference \cite{blaizot+reinosa}, a very general
property of Feynman diagrams in the imaginary-time formalism for
finite-temperature relativistic field theories (including scalar
and fermionic fields), previously put forward as a conjecture in
reference \cite{espinosa+stockmeyer}. This property states that
the full result of performing the Matsubara sum associated to any
given Feynman graph can be obtained from its zero-temperature
counterpart, by means of a simple linear operator, given by
\eqref{eq:thermal-operator-form1}, whose form depends solely on
the topology of the graph.

The thermal operator \eqref{eq:thermal-operator-form1} has the
important feature of being independent of the discrete Euclidean
energies carried by the external lines of the graph. It follows
from this that all issues related to analytic continuations of
imaginary-time formalism Green's functions \cite{Brandt
et.al.,analytic continuations,cutting rules,weldon} can be
completely settled at the zero-temperature level. The implications
of this fact as well as the connections of the thermal operator
representation with the real-time formalism are under
investigation and will be presented elsewhere.

We have not studied the validity of the thermal operator
representation in the case of gauge theories. It would be
interesting to determine the classes of gauge fixings under which
it holds, clarifying the role of ghost fields in the formalism.

\section{Acknowledgement}
The diagrams presented in this paper were produced with JaxoDraw
\cite{jaxodraw}.
This work was supported by CONICYT, under grant Fondecyt 1030363.

\newcommand{\xprd}[3]{Phys.~Rev.~{\bf D#1}, #3 (#2)}
\newcommand{\xprl}[3]{Phys.~Rev.~Lett.~{\bf #1}, #3 (#2)}
\newcommand{\xpr}[3]{Phys.~Rev.~{\bf #1}, #3 (#2)}
\newcommand{\plb}[3]{Phys.~Lett.~{\bf B#1}, #3 (#2)}
\newcommand{\pla}[3]{Phys.~Lett.~{\bf A#1}, #3 (#2)}
\newcommand{\hepth}[1]{arXiv:hep-th/#1}
\newcommand{\hepph}[1]{arXiv:hep-ph/#1}
\newcommand{\condmat}[1]{arXiv:cond-mat/#1}
\newcommand{\mpla}[3]{Mod.~Phys.~Lett.~{\bf A#1}, #3 (#2)}
\newcommand{\jhep}[3]{JHEP~{\bf #1}, #3 (#2)}
\newcommand{\xrmp}[3]{Rev.~Mod.~Phys.~{\bf #1}, #3 (#2)}
\newcommand{\jmp}[3]{Jour.~Math.~Phys.~{\bf #1}, #3 (#2)}
\newcommand{\npb}[3]{Nucl.~Phys.~{\bf B#1}, #3 (#2)}
\newcommand{\epjc}[3]{Eur.~Phys.~J.~{\bf C#1}, #3 (#2)}
\newcommand{\nuocim}[3]{Nuov.~Cim.~{\bf #1}, #3 (#2)}
\newcommand{\ptp}[3]{Prog.~Theoret.~Phys.~{\bf #1}, #3 (#2)}

\end{document}